\documentclass[conference]{IEEEtran}
\usepackage{graphicx}%------------------------------added by ZY
\usepackage{subfigure}
\usepackage{float}%---------------------------------added by ZY
\usepackage{mathrsfs}%------------------------------added by ZY
\usepackage{amsfonts}
\usepackage{array}
\usepackage{amssymb,amsmath}
\usepackage{longtable}
\usepackage{rotating}
\usepackage{multirow}
\usepackage{epstopdf}
\usepackage[lined,boxed,commentsnumbered, ruled]{algorithm2e}
\usepackage{flushend}%%%%%%%%%%%%%%%%%%%%%%%%%%%%%%%%%
\usepackage{cite}
\usepackage{float}
\usepackage{stfloats}
\usepackage[colorlinks,
            linkcolor=black,
            anchorcolor=black,
            urlcolor=blue,
            citecolor=black]{hyperref}
\usepackage[hyphenbreaks]{breakurl}

\renewcommand{\baselinestretch}{1}
\def\BibTeX{{\rm B\kern-.05em{\sc i\kern-.025em b}\kern-.08em
    T\kern-.1667em\lower.7ex\hbox{E}\kern-.125emX}}

\renewcommand{\baselinestretch}{1}
\newcommand{\methodname}{{{FedRC}}}    
\begin{document}

%\title{Federated Reservoir Computing Empowered GNSS Signals' Interference Classification}
\title{GNSS Interference Classification Using Federated Reservoir Computing }
\author{\IEEEauthorblockN{Ziqiang Ye\textsuperscript{1}, Yulan Gao\textsuperscript{2}, Xinyue Liu\textsuperscript{3}, Yue Xiao\textsuperscript{1}, Ming Xiao\textsuperscript{2}, and Saviour Zammit\textsuperscript{4} }
\IEEEauthorblockA{\textsuperscript{1}{National Key Laboratory of Wireless Communications} \\
{University of Electronic Science and Technology of China, Chengdu, 611731, China}\\
\IEEEauthorblockA{\textsuperscript{2}{Division of Information Science and Engineering, KTH Royal Institute of Technology, 100 44 Stockholm, Sweden}}
\IEEEauthorblockA{\textsuperscript{3}{SWJTU-Leeds Joint School, Southwest Jiaotong University, Chengdu, 611756, China}}
\IEEEauthorblockA{\textsuperscript{4}{Department of Communications and Computer Engineering, University of Malta, Msida MSD 2080, Malta}}
{Email: xiaoyue@uestc.edu.cn}
}
}

\maketitle

\begin{abstract} 
The expanding use of Unmanned Aerial Vehicles (UAVs) in vital areas like traffic management, surveillance, and environmental monitoring highlights the need for robust communication and navigation systems.
Particularly vulnerable are Global Navigation Satellite Systems (GNSS), which face a spectrum of interference and jamming threats that can significantly undermine their performance.
While traditional deep learning approaches are adept at mitigating these issues, they often fall short for UAV applications due to significant computational demands and the complexities of managing large, centralized datasets.
In response, this paper introduces Federated Reservoir Computing (FedRC) as a potent and efficient solution tailored to enhance interference classification in GNSS systems used by UAVs. 
Our experimental results demonstrate that FedRC not only achieves faster convergence but also sustains lower loss levels than traditional models, highlighting its exceptional adaptability and operational efficiency.
\end{abstract}


\section{Introduction}
Global Navigation Satellite Systems (GNSS) such as GPS, Galileo, GLONASS, and Beidou are pivotal for a myriad of applications, underpinning the modern economy.
These systems are indispensable for aviation, where they are utilized for navigating under instrument flight rules, especially when visual flight is not feasible, as well as for precise approaches and landings.
Additionally, GNSS plays a crucial role in maritime safety through the automatic identification system and aids collision avoidance.
Its utility spans various sectors, including agriculture and the burgeoning field of autonomous spacecraft \cite{kaplan2017understanding}.
The reliance on GNSS is so profound that a disruption lasting five days could potentially inflict an economic damage of £5.2 billion in the UK alone \cite{martins2014gnss}.
By 2020, essential services like ATMs, power grids, and railway systems were deeply integrated with GNSS technologies, with any interference posing grave national security risks \cite{sathyamoorthy2013global}.
The 5G networks are also heavily dependent on GNSS for synchronization, where interruptions could severely degrade service quality \cite{colard2020distributing,gao2018game}.

The system's architecture, encompassing satellites, control segments, monitoring stations, and user components, is inherently vulnerable to jamming due to the satellite signals' low strength, approximately $-130\hbox{ dBm}$ \cite{ioannides2016known}.
A notable incident in 2013 involved a truck driver being fined \$32,000 for operating a GNSS jammer that interfered with the  Newark Airport's Air Traffic Control System \cite{merz2013gps}.
Furthermore, over a span of two years, the European Global Navigation Satellite Systems Agency documented 450,000 jamming incidents, many of which significantly impacted GNSS operations \cite{borio2012gnss}.
These jammers, which can be purchased online for as little as £10 \cite{humphreys2017interference}, highlight the urgent necessity for effective jamming mitigation strategies. 
These strategies must begin with the robust detection and precise classification of interference signals. 

The growing need for robust GNSS services has driven significant advances in interference detection and classification technologies. 
Early classification systems identified four main types if civilian GNSS jammers \cite{kraus2011survey}; by 2019, this number had expanded to six \cite{morales2019jammer}.
Such detailed classification is crucial for the development of effective countermeasures that ensure seamless GNSS operation. 
Concurrently, research in related areas such as wireless communication and spectrum monitoring has leveraged deep learning \cite{o2017spectral, arjoune2020novel} and Convolutional Neural Networks (CNNs) \cite{swinney2021unmanned,swinney2020unmanned,jiang2024federated} to significantly improve the capabilities for signal detection and classification. 
% Class I, Amplitude Modulated (AM) jammers; Class II, Chirp jammers; Class III, Frequency Modulated (FM) jammers; Class IV, Pulse or Distance Measurement Equipment (DME) jammers; Class V, Narrowband (NB) jammers; and Class VI, scenarios with no jamming signal.
% Additionally, Ferre \emph{et al.} \cite{ferre2019image} developed and released an open-access dataset specifically for GNSS jammer classification.
% This dataset includes 61,800 spectrogram images with a $512 \times 512$ resolution, covering both clean and jammed GNSS signals across various Carrier-to-Jammer Signal Ratios.
% Tandiya \emph{et al.} leveraged spectrograms stored as 2D images with a video prediction system known as Prednet to identify jamming by comparing the spectrum with a no-jamming reference \cite{tandiya2018deep}.
% O'Shea \emph{et al.} \cite{o2017spectral} applied CNNs to spectrograms for detecting and classifying various wireless signals, including GSM, Bluetooth, and LTE, in the context of spectrum monitoring.
% Arjoune \emph{et al.} \cite{arjoune2020novel} evaluated several machine learning models, such as Support Vector Machine (SVM), Random Forest (RF), and Neural Networks, for detecting the presence of jamming signals in wireless communications.
% However, their study focused solely on detection, with classification suggested as a direction for future research.
% Jiang \emph{et al.} \cite{} explored the use of the VGG16 CNN model for signal classification.

Despite these advancements, the challenges of real-world data collection and the high computational demands of traditional deep learning methods on mobile platforms, such as Unmanned Aerial Vehicles (UAVs), necessitate innovative approaches like reservoir computing. 
This technique not only overcomes computational limitations but also aligns federated learning models, which enhance data privacy and reduce network load by processing data locally \cite{niknam2020federated}.

In this context, this paper set out to harness federated reservoir computing (FedRC) to develop a distributed interference signal classification system that is both efficient and privacy-preserving.
By leveraging local model training and aggregation without direct data sharing, our proposed method strikes a balance between privacy protection and computational efficiency for UAVs tasked with GNSS integrity operations. 
This approach is particularly suited to environments where traditional data collection and processing methods are unfeasible, thus ensuring robust GNSS functionality amidst diverse and evolving threats.
Extensive comparative experiments based on non-IID data distributions demonstrate that our proposed algorithm achieves $38.5\%$ lower loss and $31.67\%$ lower running time on average than the best-performing baseline.

\section{System Model}
\subsection{System Overview}
For the purpose of this article, we model the analog baseband equivalent of the received GNSS signal as follows:
\begin{align} \label{eq:1}
    r(t) = s(t)+j(t)+w(t),
\end{align}
where $s(t)$ denotes the desired GNSS satellite signals, and $w(t)$ accounts for random elements such as thermal noise, typically characterized as an additive white Gaussian noise (AWGN) process.
Inspired by \cite{morales2019survey}, the term $j(t)$ refers to the interference signal waveform as detected at the receiver, varying according to the type of jammer.
Accurate identification of $j(t)$ is crucial for swiftly countering jamming threats through either localization  or mitigation strategies \cite{nardin2023crowdsourced}.
In this context, Interference Classification (IC) techniques are employed to deduce the waveform of $j(t)$, which enables its direct subtraction from $r(t)$.
As aforementioned, accurately identifying $j(t)$'s waveform is instrumental for various applications, including its mitigation and reconstruction using IC methods.

According to \cite{morales2019jammer},  Wide Band (WB) jammers are not included in our analysis due to the difficulties associated with their detection in spectrogram analyses. 
In contrast, Amplitude Modulated (AM) and Frequency Modulated (FM) jammers, which are characterized by narrow spectral signatures, often overshadow the signal of interest, making it indistinguishable with noise. 
Our classfication strategy, detailed in Section III, effectively identifies the presence or absence of such interference. 
The specific waveform patterns of interference, denoted as $j(t)$, are categorized in \cite{morales2019jammer} as follows: Class I includes AM jammers; Class II consists of Chirp jammers; Class III covers FM jammers; Class IV comprises Pulse or Distance Measurement Equipment (DME) jammers; Class V encompasses Narrowband (NB) jammers; and Class VI identifies scenarios with no jamming signal.

\begin{figure}[!t]
	\centering
	\includegraphics[width=\linewidth]{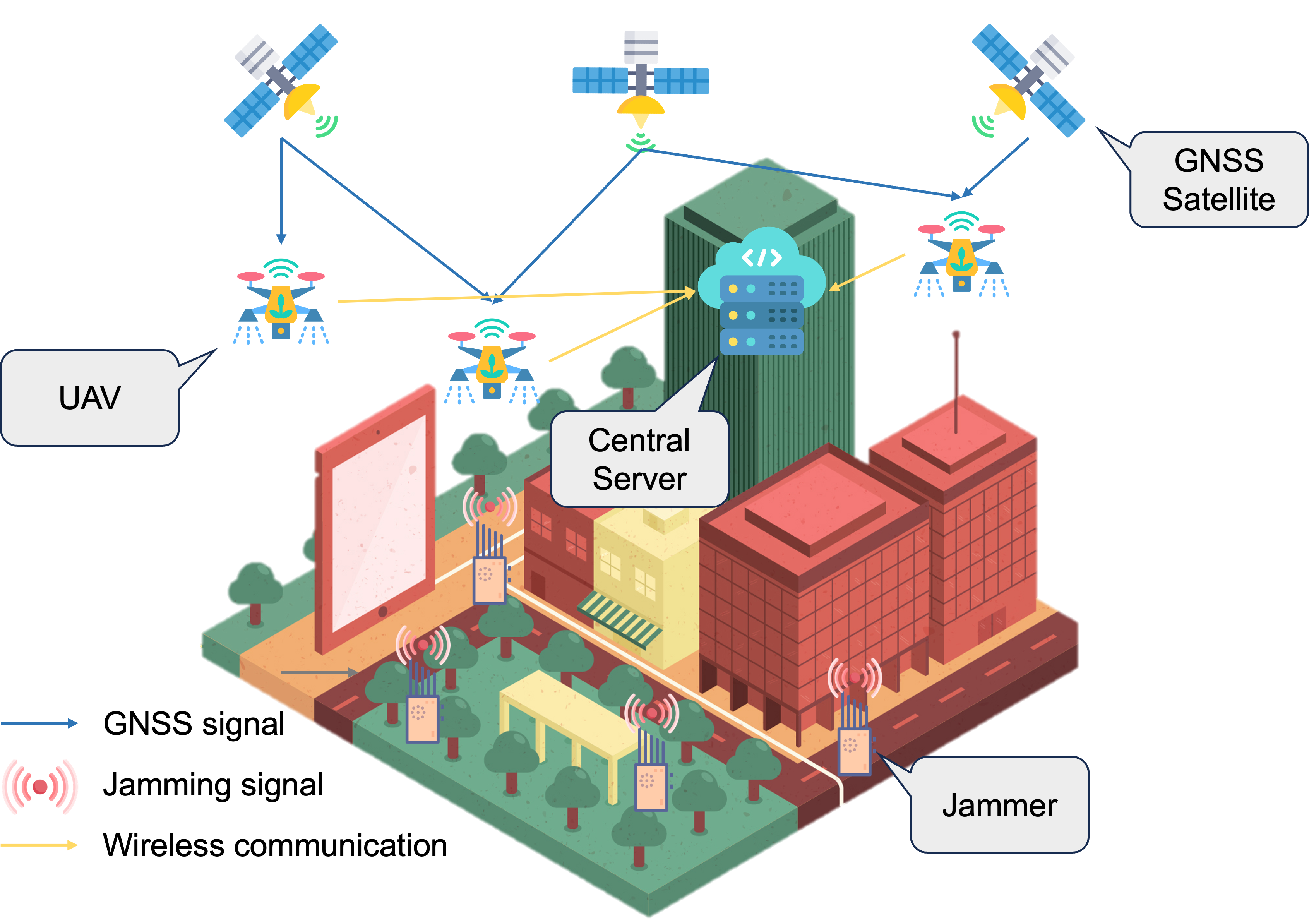}
	\caption{Architectural diagram of UAV-Based GNSS Interference Classicication utilizing federated reservoir computing.}
	\label{fig:1}
\end{figure}

In the system configuration depicted in Fig.\ref{fig:1}, a group of UAVs, collectively denoted as $\mathcal{U}$, are distributed randomly across a circular area and maintain a consistent velocity throughout each time slot.
These UAVs, tasked with receiving GNSS signals, must operate within their computational limits, often making it unfeasible to run complex models onboard.
Concurrently, an assortment of jammers $j \in \mathcal{J}$, also randomly positioned within the area, engage in various interference activities as described in \cite{morales2019jammer}, randomly choosing their interference modes with each time slot.

To combat these disruptions, each UAV is equipped with a FedRC model.
This setup allows the UAVs to partake in distributed training activities, collectively improving their capability to recognize interference without necessitating the transmission of substantial data volumes to a central server.
After local training, the UAVs relay model parameter updates to a central server, which then consolidates these updates to refine the global model.
This model is subsequently redistributed to the UAVs.
This cycle of local computation and central aggregation ensures that the UAVs efficiently use their computational resources while effectively detecting interference throughout the network.

% \section{Problem Formulation}
% In this section, 

\section{Federated Reservoir Computing Design}
Echo State Networks (ESN) present distinctive training characteristics that substantially enhance federated learning approaches.
The principal advantage of using ESN is that the collective model closely emulates what would be attained if all input data were centrally compiled and processed.
In the framework we have developed, named \methodname{}, we decompose the standard readout training equation algebraically, represented as follows:
\begin{align}
\mathbf{\Theta}=\mathbf{Y}\mathbf{\Phi }^T(\mathbf{\Phi }\mathbf{\Phi }^T+\beta \mathbf{I})^{-1},
\end{align}
where $\mathbf{\Theta}$ represents the readout weight matrix, $\mathbf{Y}$ is the target matrix, and $\mathbf{\Phi}$ aggregates the collected input data matrix.
The term $\beta$ signifies the $L_2$ regularization factor, selected through model optimization.

Under the Federated Averaging scheme, we ensure uniformity in the reservoir configurations across all UAVs.
Each UAV $u$ computes the matrices $\mathbf{\Gamma}$ and $\mathbf{\Omega}$ as outlined below:
\begin{align} \label{eq:2}
    & \mathbf{\Gamma}_u = \mathbf{Y}_u\mathbf{\Phi }_u^T, \\ \notag
    & \mathbf{\Omega}_u = \mathbf{\Phi }_u\mathbf{\Phi }_u^T + \beta_u \mathbf{I}.
\end{align}
The matrices $\mathbf{\Gamma}_u$ and $\mathbf{\Omega}_u$ are transmitted to the server, where they are aggregated according to the equations below:
\begin{align} \label{eq:3}
    & \mathbf{\Gamma} = \sum_{u \in \mathcal{U}}\mathbf{\Gamma}_u, \\ \notag
    & \mathbf{\Omega} = \sum_{u \in \mathcal{U}}\mathbf{\Omega}_u.
\end{align}
Upon aggregating the matrices as detailed in Eq.\eqref{eq:3}, the server then calculates the optimal readout weights
\begin{align} \label{eq:4}
    \mathbf{\Theta} = \mathbf{\Gamma}\mathbf{\Omega}^{-1}.
\end{align}
It is important to note that Eq. \eqref{eq:4} is mathematically identical to Eq. \eqref{eq:1} if the server had access to all the data locally.
Once the optimal weights are computed as per Eq. \eqref{eq:4}, they are redistributed back to the UAVs following the FedAvg protocol.

The superiority of our approach lies in its capability to perform efficient one-shot training across potentially limitless data volumes.
This efficiency arises because the matrices $\mathbf{\Gamma}$ and $\mathbf{\Omega}$ do not depend on the number of training sequences.
When new data becomes available, these matrices can be incrementally updated by the clients by simply adding the new computational results.
For instance, if UAV $v$ has previously computed the matrices $\mathbf{\Gamma}_u$ and $\mathbf{\Omega}_u$ over $t$ iterations, the incorporation of new data and corresponding labels allows the client to calculate the updated matrices $\tilde{\mathbf{\Gamma}}_u$ and $\tilde{\mathbf{\Omega}}_u$, based exclusively on this newly available information.
These updates are then aggregated as follows:
\begin{align} \label{eq:5}
    \mathbf{\Gamma}_u^{t+1} = \mathbf{\Gamma}_u^{t} + \tilde{\mathbf{\Gamma}}_u, \\ \notag
    \mathbf{\Omega}_u^{t+1} = \mathbf{\Omega}_u^{t} + \tilde{\mathbf{\Omega}}_u. 
\end{align}
This structured approach ensures that our federated learning system remains scalable, adaptable, and highly effective across diverse and dynamically changing data environments.

\section{Simulation Results}
We assess the efficacy of our \methodname{}.
This approach enables a jammer classifier, operating in a distributed framework, to approximate the performance levels of a system trained centrally with access to comprehensive local datasets.
Initially, we outline the dataset employed; next, we detail its use within the distributed learning architecture, describe the configuration of the model, and ultimately, we present the results obtained.

\subsection{Data Preprocessing}
For the experiments outlined here, we employed the dataset sourced from \cite{ferre2019image}.
In an effort to optimize the training procedure and efficiently utilize computational resources, we applied several preprocessing techniques commonly used in machine learning.
In this instance, we merged the training and validation datasets—usually reserved for hyperparameter adjustments—and divided the combined data into an $80\%$ training and $20\%$ testing split.
Additionally, we enhanced the dataset quality by downsizing the image resolution from $512 \times 512$ to $256 \times 256$ pixels through bilinear interpolation.
Following these modifications, the data was normalized and transformed into 1D time series, streamlining the subsequent training process.

\begin{figure}[!t]
	\centering
	\includegraphics[width=\linewidth]{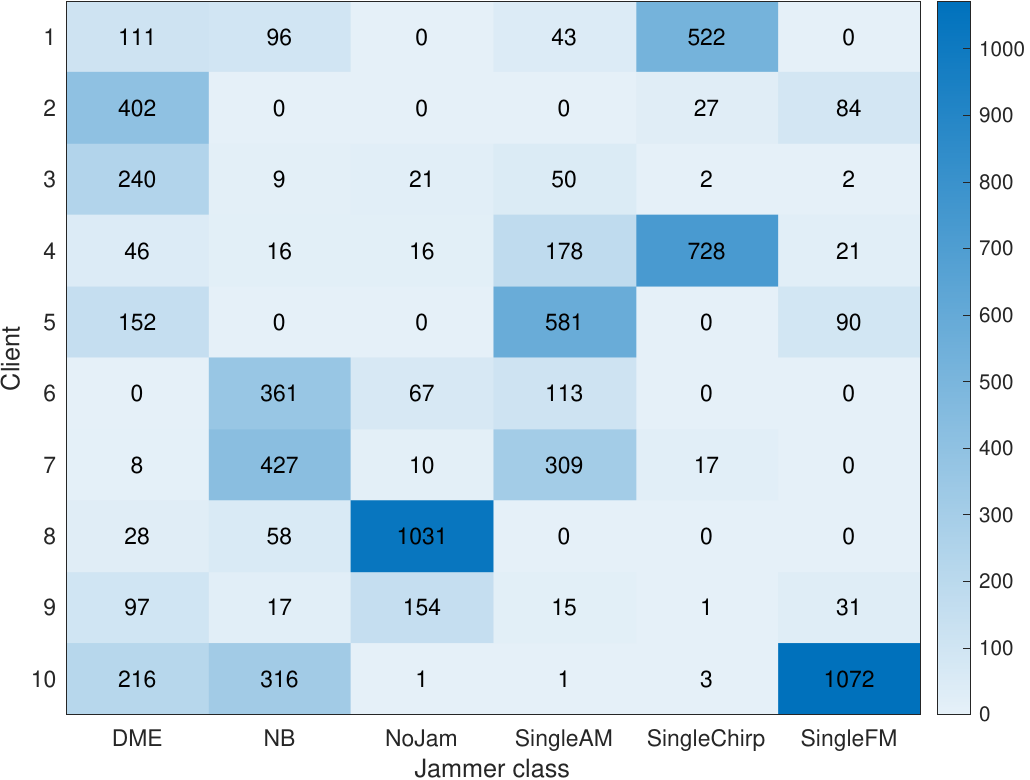}
	\caption{Data distribution.}
	\label{fig:2}
\end{figure}

\subsection{Federated Data Setting}
We explored two distinct data configurations in our study.
The initial scenario involved an IID setting where each client received a comparable distribution of data, specifically an equal number of samples from each class.
Here, the data was evenly distributed among 10 clients, giving roughly 500 samples per client.

The alternative scenario focused on a non-IID setting, characterized by an uneven distribution of class labels across the training data.
For creating these non-IID divisions, we utilized a method detailed in \cite{li2020federated}, employing a Dirichlet distribution to sample the client data.
The concentration parameter was set at a relatively low value of $0.1$. The effects of this setting are illustrated in Fig.\ref{fig:2}, which depicts the variation in the number of samples per class for each UAV when $U=10$.
This arrangement resulted in a disparate distribution of data, where certain clients ended up with either significantly more or fewer samples of specific class labels.

\subsection{Model Setting}
In this section, we report the hyperparameters for the \methodname{}.
The units in the \methodname{} is set to $500$, the spectral radius is set to $0.1$. The leaking rate and input scaling is set to $0.1$ and $0.1$, respectively.
The input connectivity and recurrent connectivity is set to $3$ and $8$, separately.

For the comparison, we employ a CNN network which consists of three convolutional layers, each followed by a ReLU activation function and max-pooling.
Additionally, a fully connected deep neural network comprising six linear layers is also employed for comparison.

In federated process, the local batchsize is set to $256$, local training epoch is $10$.
The fraction of number of UAV participated in the FL process is $0.9$.
We assume that there are $10$ UAVs and $6$ jammers in the system.

\begin{figure}[!t]
	\centering
	\includegraphics[width=\linewidth]{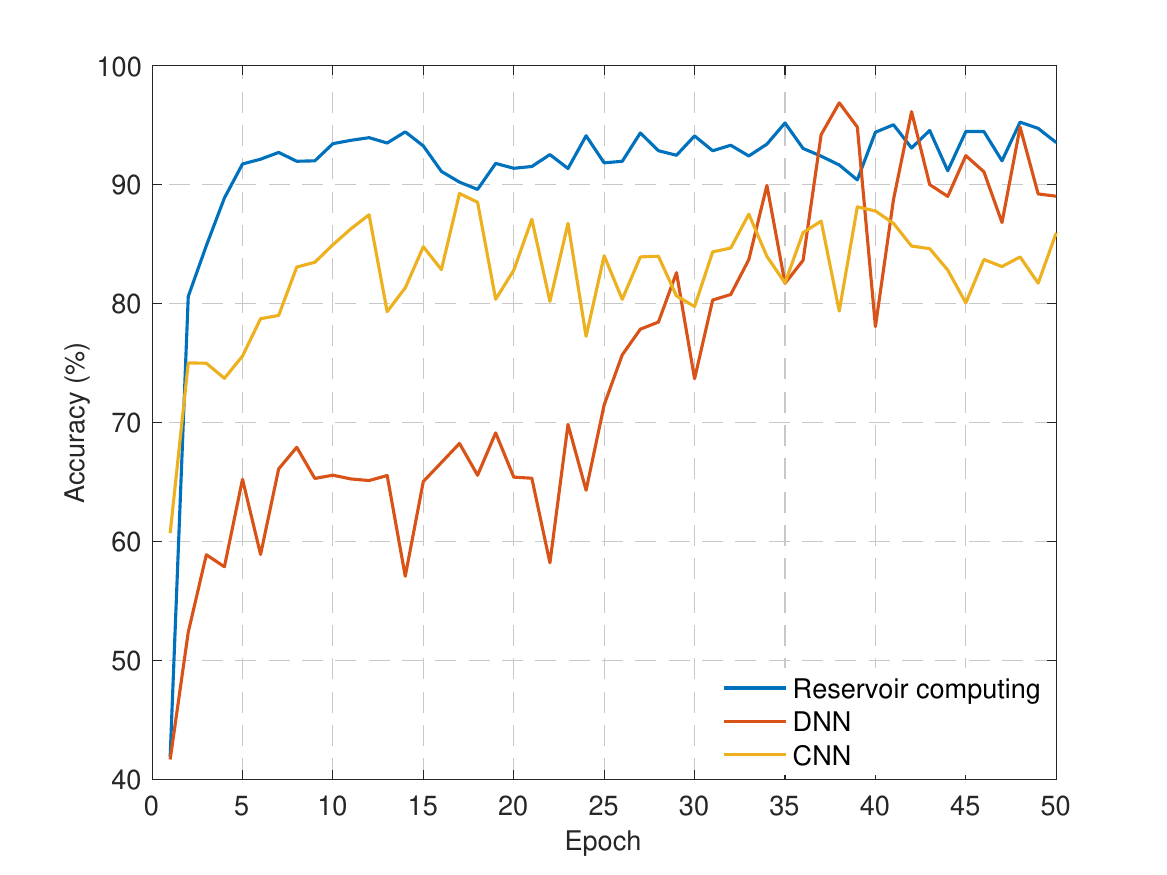}
	\caption{Comparison of Classification Accuracy Across Different Models.}
	\label{fig:3}
\end{figure}

\subsection{Results Analysis}
During a comprehensive 50-round training iteration involving full dataset processing, it was observed that Reservoir Computing required only 2174 seconds, while the CNN and DNN networks took 3182 seconds and 3140 seconds, respectively.
This finding highlights the superior efficiency of Reservoir Computing in comparison to the traditional deep learning networks.
Specifically, Reservoir Computing not only demonstrates a more efficient processing time but also significantly outperforms CNN and DNN in terms of training speed.

The comparison of classification accuracy across different models, as illustrated in Fig. \ref{fig:3}, demonstrates that the reservoir computing model consistently outperforms both the DNN and CNN models over the course of 50 epochs.
The reservoir computing model achieves a higher accuracy early in the training process and maintains a more stable performance, with accuracy consistently above 85\% after the initial epochs.
In contrast, the DNN and CNN models exhibit greater fluctuations in accuracy, indicating potential instability and overfitting, particularly in the later stages of training.
The CNN model, while showing some promise in the early epochs, fails to sustain its performance, with accuracy eventually falling behind that of the reservoir computing model.
This comparison highlights the efficacy of reservoir computing in environments where computational resources are limited, such as UAVs, due to its lower complexity and higher robustness in maintaining classification accuracy.

\begin{figure}[!t]
	\centering
	\includegraphics[width=\linewidth]{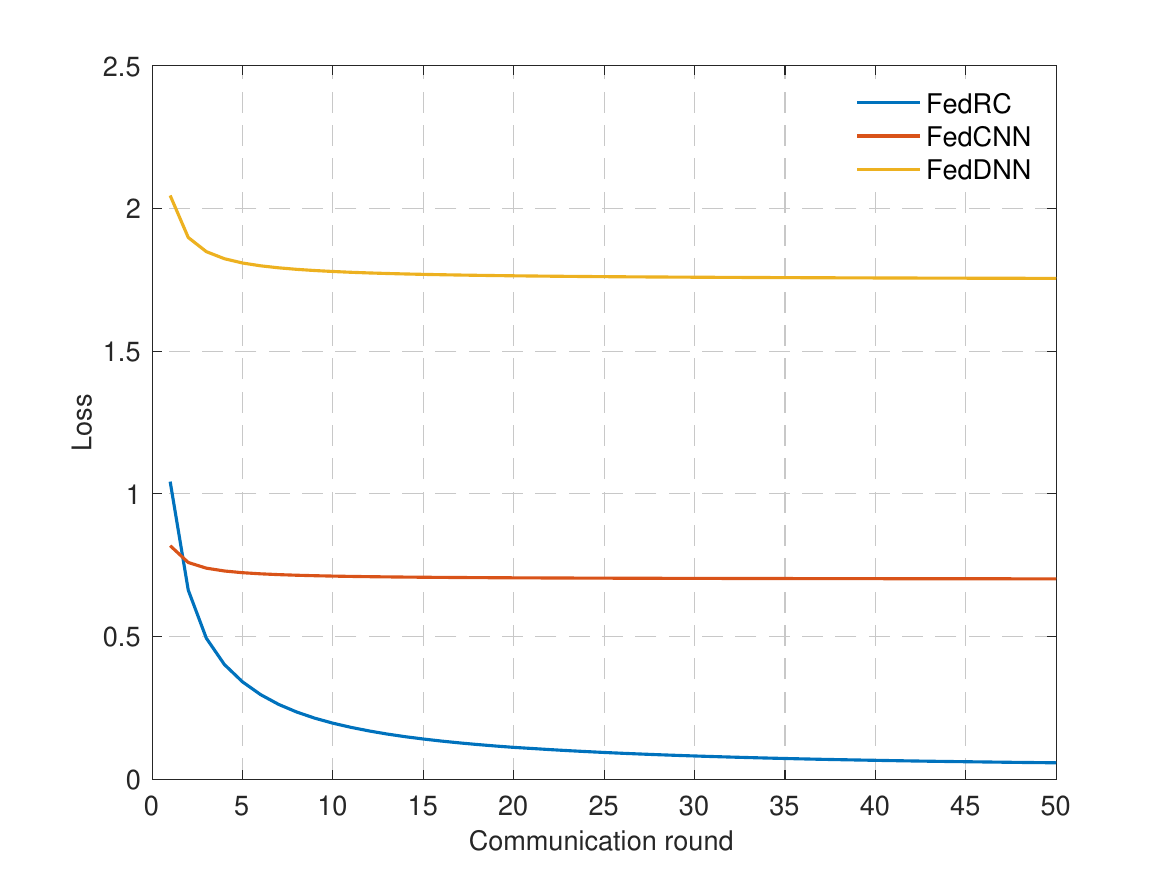}
	\caption{Loss in 50 rounds under IID data setting.}
	\label{fig:4}
\end{figure}

The graph depicted in Fig. \ref{fig:4} illustrates the loss metrics over 50 communication rounds under an IID data setting for three different federated learning models: FedRC, FedCNN, and FedDNN.
It is evident that FedRC shows a significant reduction in loss more rapidly compared to the other models, stabilizing at a lower value early in the training rounds.
In contrast, FedCNN and FedDNN exhibit higher loss values throughout the rounds, with FedCNN gradually approaching a plateau slightly above 1, and FedDNN maintaining a consistent loss close to 2.
This suggests that FedRC is not only more efficient in terms of convergence speed but also achieves a more optimal performance in handling IID data settings.
The performance disparity indicates potential differences in the models' sensitivity to initial conditions and their capacity to generalize from the distributed data.

\begin{figure}[!t]
	\centering
	\includegraphics[width=\linewidth]{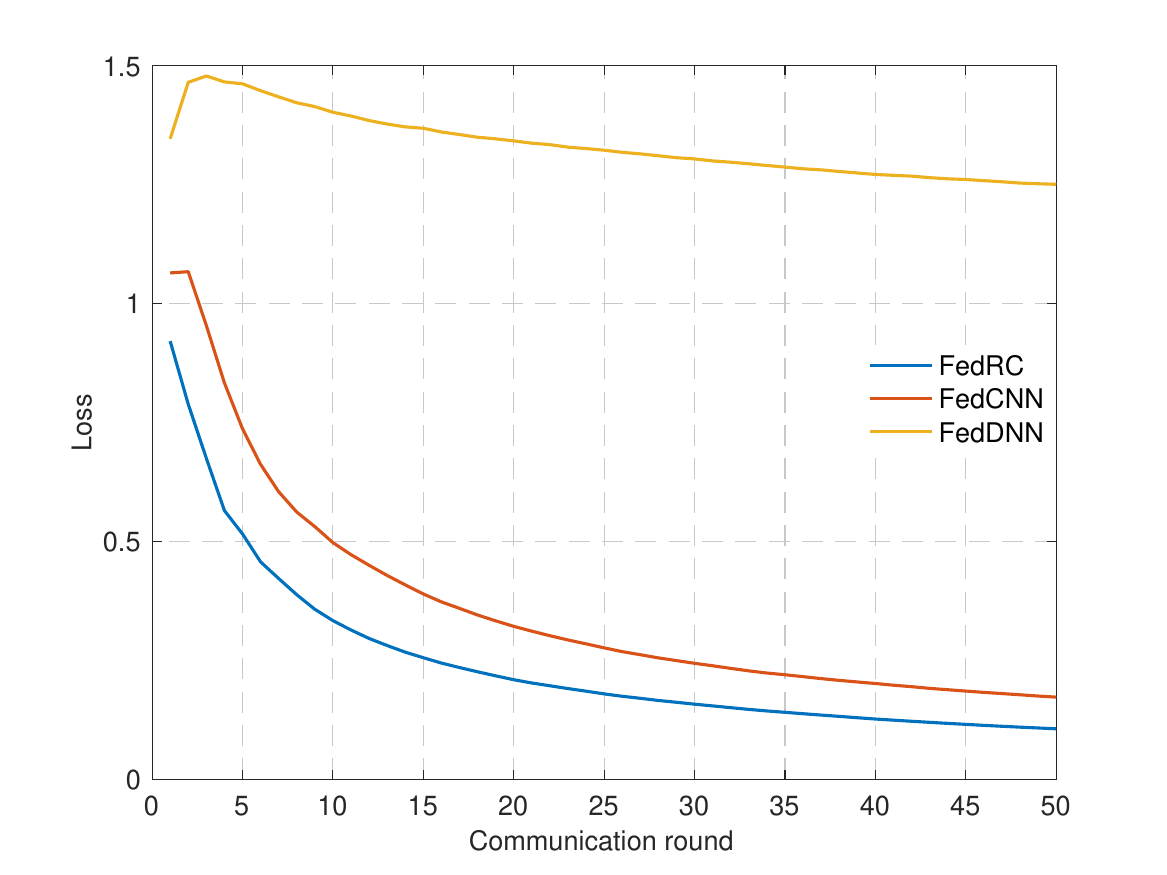}
	\caption{Loss in 50 rounds under Non-IID data setting.}
	\label{fig:5}
\end{figure}

Fig. \ref{fig:5} illustrates the performance of three federated learning models—FedRC, FedCNN, and FedDNN—across 50 communication rounds in a non-IID data setting.
The graph demonstrates that FedRC exhibits the steepest decline in loss, indicating rapid convergence compared to the other models.
By around the $10$-th round, FedRC's loss stabilizes near a value of $0.5$, maintaining this low level throughout the remaining rounds.
In contrast, FedCNN and FedDNN show a slower descent in loss.
FedCNN begins with a higher loss, crossing below the $1.0$ mark around the $15$-th round and continuing to decrease gradually, but never matching the low level of FedRC.
Meanwhile, FedDNN starts with the highest loss and experiences a moderate reduction, stabilizing just below $1.5$, which suggests it struggles more significantly with the non-IID data distribution.
\section{Conclusion}
This study has successfully demonstrated that FedRC offers significant advantages over traditional federated deep learning models.
The experiments conducted show that FedRC not only achieves faster convergence and maintains lower loss in non-IID data settings but also significantly reduces computational and communication overheads.
These attributes make FedRC highly suitable for deployment in distributed systems where real-time data processing and privacy are paramount.
By leveraging the lightweight and adaptable nature of FedRC, our findings suggest that UAVs equipped with this technology can effectively detect and mitigate GNSS jamming, enhancing their reliability and operational efficiency in critical applications.
% \section*{Acknowledgement}
% This work is 
\bibliographystyle{ieeetr}
\bibliography{conference_101719}

\end{document}